\documentclass{emulateapj}

\slugcomment{Submitted to ApJL}

\shorttitle{The peculiar solar composition}
\shortauthors{Mel\'endez et al.}

\begin{document}


\title{The peculiar solar composition and its possible relation to planet formation}


\author{J. Mel\'{e}ndez\altaffilmark{1}}
\affil{Centro de Astrof\'{i}sica da Universidade do Porto, Rua das Estrelas, 4150-762 Porto, Portugal}
\email{jorge@astro.up.pt}
\author{M. Asplund}
\affil{Max-Planck-Institut f\"ur Astrophysik, Karl-Schwarzschild-Str. 1, Postfach 1317, D-85741 Garching, Germany}
\author{B. Gustafsson}
\affil{Institutionen f\"or fysik och astronomi, Uppsala universitet, Box 515, SE-75120 Uppsala, Sweden}
\and
\author{D. Yong}
\affil{Research School of Astronomy \& Astrophysics, Australian National University, Mount Stromlo Observatory, Cotter Road, Weston Creek, ACT 2611, Australia}


\altaffiltext{1}{Visiting Astronomer, Magellan Telescopes, Las Campanas Observatory, Chile,
and W.M. Keck Observatory, Manua Kea, Hawaii}


\begin{abstract}
We have conducted a differential elemental abundance analysis of unprecedented 
accuracy ($\sim0.01$ dex)
of the Sun relative to 11 solar twins from the Hipparcos catalogue and 
10 solar analogs from planet searches. 
We find that the Sun shows a characteristic signature with a 
%%MA
$\approx 20$\% 
depletion 
of refractory elements relative to the volatile elements in comparison with the 
solar twins.
The abundance differences correlate strongly with the condensation temperatures 
of the elements. 
This peculiarity also holds in comparisons with solar analogs 
known to have close-in giant planets while the majority of solar analogs
found not to have such giant planets in radial velocity 
monitoring show the solar abundance pattern. We discuss 
various explanations for this peculiarity, including the possibility that 
the differences in abundance patterns are related 
to the formation of planetary systems like our own,
%%MA
in particular to the existence of terrestrial planets.
\end{abstract}





\section{Introduction}

Studies of extra-solar planetary systems and the possibility of life in the Universe 
depend fundamentally on whether the Sun and its planets are typical 
\citep{gus08, bee04, rob08, seg03}. Previous chemical composition studies 
have concluded that the Sun is a typical star \citep{gus98, gus08, rob08}.  
Besides an increased likelihood for stars with higher overall metallicity of hosting 
giant planets, no definitive chemical differences have been established between 
stars with and without known planets \citep{gon97, ecu06, udr07}.
Previous studies on chemical abundance anomalies in the Sun as compared 
with solar-like stars have been inconclusive due to the 
relatively large 
%%MA 
($\ga 0.05$ dex) 
remaining systematic errors \citep{gus08, rob08, red03}. 
In such studies, the selection of stars to which the Sun is compared is crucial. 
Different categories of solar-like stars
can be identified. ``Solar-type stars'' range from late F to early K, 
``solar analogs'' include only G0-G5 dwarfs, 
and stars almost identical to the Sun are called ``solar twins'' \citep{cay96}. 
In modern stellar chemical analyses, the error
budget is dominated by systematic errors in the model atmospheres and 
modeling of stellar spectra \citep{asp05}.  
The highest possible accuracy in the analysis of the Sun relative to stars can be 
achieved in comparisons relative to solar twins, since the model errors  
will cancel to a large extent and effects of possible systematic errors 
in temperature scale and absolute abundances are minimized. 
 
 \section{Observations and analysis}

%% In a manner similar to \objectname authors can provide links to dataset
%% hosted at participating data centers via the \dataset{} command.  The
%% second curly bracket argument is printed in the text while the first
%% parentheses argument serves as the valid data set identifier.  Large
%% lists of data set are best provided in a table (see Table 3 for an example).
%% Valid data set identifiers should be obtained from the data center that
%% is currently hosting the data.
%%
%% Note that AASTeX interprets everything between the curly braces in the 
%% macro as regular text, so any special characters, e.g. "#" or "_," must be 
%% preceded by a backslash. Otherwise, you will get a LaTeX error when you 
%% compile your manuscript.  Special characters do not 
%% need to be escaped in the optional, square-bracket argument.

Our sample comprises 11 solar twins and 10 solar analogs.
The solar twins were selected from more than 100,000 
stars in the Hipparcos catalogue by employing precise colour-temperature relations, 
trigonometric parallaxes and age indicators \citep{mel06, mel07}.  
They lie within a distance of 75 pc from the Sun and belong to 
the Galactic thin disk population. No giant planets are known around
our solar twins because most of them have not been searched for planets.
The solar analogs were selected from radial velocity planet surveys \citep{udr07}.
The Sun (reflected light from the asteroid Vesta), solar analogs and 
solar twins were observed with the MIKE spectrometer 
at the Clay 6.5 m Magellan telescope at Las Campanas Observatory in Chile, 
at high spectral resolution ($R = \Delta\lambda/\lambda$ = 65, 000) and 
very high signal-to-noise ratio ($S/N \sim 450$ per pixel at 600 nm) through 
the wavelength range 340-1000 nm. 
Example Magellan spectra of solar twins are presented in Fig. 1. 
Finally, one of our twins (HIP79672=18 Sco) was also observed with the 
HIRES spectrograph on the Keck I telescope to yield an even higher quality 
spectrum with $S/N\approx 450$ and $R=100,000$;  a solar spectrum
(asteroid Ceres) was also obtained with identical setup.

%% In this section, we use  the \subsection command to set off
%% a subsection.  \footnote is used to insert a footnote to the text.


%% This section also includes several of the displayed math environments
%% mentioned in the Author Guide.

A model independent analysis \citep{mel06, mel07} indicates that the twins 
have effective temperatures within 75 K of the Sun, 
logarithmic surface gravities within 0.10 dex and logarithmic iron abundances [Fe/H] 
within 0.07 dex (where 
%%MA
[A/B]$=log_{10}(N_A/N_B)_{star}-log_{10}(N_A/N_B)_{\odot})$. 
These indications were confirmed by a detailed model-atmosphere analysis.
The final adopted effective temperatures
were obtained from the excitation equilibrium of Fe I, while the surface gravities 
were determined from the ionization equilibrium of Fe I/Fe II.
We have achieved an accuracy in [X/Fe] at the level of 0.01 dex
for the solar twins (e.g., the star-to-star scatter 
in [Cr/Fe] is only $\sigma = 0.009$ dex). 
For the solar analogs high precision abundances are also obtained due to 
their similarity to the Sun, although not as high as for the solar twins.
%%MA ($\approx 0.02$\,dex?). 
The observed scatter in relative abundances is fully consistent with the 
predicted error from the quality of the spectra and the remaining 
uncertainties in the stellar parameters.

\section{Results}
Our results reveal that the solar chemical abundances relative to iron depart from 
the mean abundance ratios in the solar twins (Fig. 2). The Sun is enriched 
by $\sim0.05$ dex in most of the lighter elements, while other elements like 
aluminium are deficient by $\sim0.03$ dex, thus the offset between the elements 
that are depleted and enhanced is $\sim0.08$ dex (20\%). 
In Fig. 3 the abundance differences are shown versus the 
50\% condensation temperature $T_{\rm cond}$  \citep{lod03} for a solar-system 
composition gas. Volatile elements (low $T_{\rm cond}$) 
are more abundant in the Sun relative to the twins while elements that easily form 
dust (elements with high $T_{\rm cond}$, i.e. refractories) are under-abundant. 
A very similar pattern, albeit of much greater amplitude, is seen in the 
interstellar medium due to dust condensation \citep{sav96}, as well as 
in $\lambda$ Bootis stars and some highly evolved stars through the 
accretion of dust-cleansed gas \citep{ven90}. 
The correlation with $T_{\rm cond}$ shown in Fig. 3 is highly significant: 
the Spearman correlation coefficient is $r_{\rm S}=-0.91$ and
an arrangement of this type have a probability of $\sim10^{-9}$ to happen by chance.
The scatter around the mean trend in Fig. 3 is only 0.01 dex. 
On a star-by-star basis, only 1-3 solar twins resemble the Sun chemically (Fig. 4). 
In fact, the probability of obtaining the same order between abundances 
differences for the four degrees of volatility for 9 (and very nearly 10) out of 11 stars 
as a result of random errors is only $2\times10^{-11}$. 
Previous attempts to trace $T_{\rm cond}$-dependent abundance differences for 
stars known to host giant planets have not convincingly demonstrated 
significant differences \citep{udr07}, probably due to the high accuracy needed.

%%MA
The results for the solar analogs are shown in Fig. 5. 
The solar analogs with detected giant planets have an 
abundance pattern similar to the solar twins,
while the solar analogs without detected giant planets 
are more similar to the Sun.

In order to confirm that our results are robust, we have checked our 
abundance scale using spectra obtained with another telescope and spectrograph. 
We repeated our highly differential abundance analysis for the Sun and 
the solar twin HIP 79672, using a HIRES/Keck spectrum. 
Our Keck-based results confirm the separation between volatile and 
refractory elements found with the MIKE data. 
The [X/Fe] abundance ratios of the HIRES analysis are in 
excellent agreement with our MIKE results. 
In particular, the relative abundances of volatiles like carbon, oxygen 
and zinc, agree to within 0.003 dex between the HIRES and MIKE data. 
This demonstrates that our abundance scale is robust, and that the quality of 
the MIKE spectra is high enough to allow our 
conclusion that the refractories in the Sun are depleted with 
respect to the volatiles.

We are not aware of any systematic errors that would, if corrected for, 
significantly reduce these differences between the Sun and the twins. 
An error in the temperature scale would lead to even larger 
departures in [X/Fe] but not generate the consistent pattern found here.

%%MA
%Below we discuss the peculiar solar composition with
%respect to the solar twins ($\S$4a-d) and the solar analogs ($\S$4e).

\section{The origin of the solar departures}

In spite of the suggestions above that the solar characteristics 
could be related to dust condensation, we shall here first discuss other
possible explanations. 

\noindent {\it(a) Galactic-evolution effects:} Perhaps the solar abundance pattern 
reflects the Galactic chemical evolution and the varying composition 
in the Milky Way ISM when the Sun was formed. 
The [$\alpha$/Fe] ratio ($\alpha$ representing O, Ne, Mg, Si, Ca, S, Ti) 
was higher earlier in the history of the Galactic disk, 
due to the 
increasing role of SN Ia relative to 
that of SN II in enriching the ISM. 
The solar twins have iron abundances similar to solar. 
If the metallicity of the Sun is unusually high for its age, 
most twins could  be younger and have lower $[\alpha/Fe]$.
The Sun is, however, not so metal-rich for its age \citep{fuh04, hol09}
and it is unlikely that the variation of [$\alpha$/Fe] with metallicity is 
large enough for this hypothesis to work \citep{ben06, che02, nis04}. 
Besides, the abundances relative to Fe for the volatile elements 
C, S and Zn as well as the refractory element Si all show no trend with metal 
content around solar metallicity. 
One direct way of testing this hypothesis would be to compare the ages of the 
twins to the Sun. Available methods for estimating stellar ages 
\citep{mel06, mel07} applied to the twins
reveals that their median age is 4.1 Gyr. 
Thus, the peculiar abundance pattern of the Sun is not the result of large age
differences between the Sun and the twins.

\citet{wie96} suggested that the Sun may have migrated from an inner
Galactic orbit. If there were radial abundance gradients,
systematic differences between the Sun and the 
twins may arise. 
However, the observed gradients in [O/Fe] for young thin disk objects are 
not steep enough \citep{prz08},
and Galactic chemical evolution models \citep{chi01} do not suggest any 
substantially steeper gradient 5 Gyr ago. 

\noindent {\it(b) Supernova pollution:}
Another possibility is that the proto-solar cloud happened 
to be more polluted by a local supernova than other star formation 
regions in the Galaxy, although this hypothesis does not explain why the Sun, 
and not the twins, happens to be affected. 
However, the peculiar solar abundance pattern cannot be explained 
by an offset of SNII relative to SNIa yields
\citep{woo95,  thi02}. 
This argument also negates the solar migration scenario. 
Finally, a suggestion that the presence of $^{60}$Fe in 
meteorites proves that the proto-solar nebula was affected by a 
nearby supernova explosion was recently found less likely \citep{wil09}.

\noindent {\it(c) Early dust separation}.
One possible explanation is that the Sun was formed in an 
interstellar cloud where dust had been blown away beforehand, 
e.g. by luminous massive stars. Another possibility could be that 
the late formation process of the Sun included accretion of gas that 
was metal-poor due to dust separation by radiation pressure, 
although it is questionable whether the accreting proto-Sun was 
efficient enough in this respect. 
Neither of these scenarios can explain why this happened to the Sun but not the twins
nor can it explain the correlation with presence of close-in giant planets ($\S$4e). 

\noindent {\it(d) Dust separation during the formation of terrestrial planets:}
A fascinating possibility is that the composition departures of the Sun 
are related to its properties as a planet host. 
A particularly striking circumstance is that the inner solar system planets 
and meteorites are enriched in refractories compared to volatiles \citep{pal00}. 
The abundance pattern of these bodies is almost a mirror-image of the solar 
pattern relative to the solar twins \citep{cie08, ale01}. 
Also, a radial gradient exists, with greater enhancement of refractories at 
smaller heliocentric distances \citep{pal00, bot06}. 
The break in the chemical abundance pattern at $T_{\rm cond} \approx 1200$ K (Fig . 3), 
suggests that the volatiles retained their original abundances both in the Sun 
and the solar twins. Such temperatures are only encountered in the inner parts ($<3$ AU) 
of proto-planetary disks \citep{cie08}, which also suggests that the abundance 
pattern is related to the presence of terrestrial planets. 

The amount of dust-depleted gas required to explain the solar abundances 
depends sensitively on the timing of the accretion. 
The mass fraction of the solar convection zone dropped from 100\% to 38\% in 10 Myr, 
and to 2\% after another 20 Myr, thereafter remaining largely constant \citep{dan94}. 
Assuming that the depletion signature was imprinted once the convection zone 
reached its present size, the removal of $\sim2\times 10^{28}$ g of refractories 
from the accreting gas would be required in order to accomplish the observed 
$\sim10\%$ reduction of refractory elements. 
This value is similar to the combined mass of refractories locked up in 
Mercury, Venus, Earth and Mars, which is $\sim 8 \times10^{27} $g  $(1.3 M_{\earth})$. 
The formation of large ($\sim 10$ km) bodies and embryos of terrestrial planets 
in the proto-solar disk is estimated to have an efficiency of 30-50\% \citep{ale01}. 
Thus, it is tempting to speculate that the formation of the terrestrial 
planets might have given the Sun its special surface composition. 
The disk masses during the T Tauri phase are $\sim 0.02$ $M_{\odot}$ but 
values up to $1 M_{\odot}$ are possible \citep{bec90, cal00}, thus enough material 
would be available to change the solar photospheric composition. 
However there is a problem with time-scales since
proto-planetary disks are observed to have typical 
life-times $<10$ Myr \citep{cal00, sic06}.
Yet, gas accretion onto 10-25 Myr old stars has been detected but is rare: 
only ~1\% of stars at an age of 13 Myr show signs of accretion \citep{cur07, whi05}. 
One possibility to explain the chemical differences with the solar twins is 
that they did not form terrestrial planets or, if they did, the 
subsequent gas accretion ocurred when the stellar convection zone 
was still deep and the planetary signatures 
thus washed out, while the Sun retained its planetary disk longer. 
A less likely explanation is that all solar twins formed terrestrial planets 
and like the Sun lost their gas disks when their convection zones were deep, 
but accreted their terrestrial planets later (after 30 Myr) when the convection 
zones were thin, thus enriching the solar twins in refractory elements. 
There are reasons to believe that thermohaline convection, 
generated by the inverse $\mu$-gradient, would then have quickly erased 
the higher metallicity of the stellar convection zone \citep{vau04}. 

\noindent {\it(e) The role of giant planets:}
The abundance pattern could also be the result of the formation of giant planets. 
The cores of Jupiter, Saturn, Uranus and Neptune containing heavy metals in the form
of rocks and ices are estimated to have a combined mass of 
$\sim30M_\earth$ \citep{gui05}. 
Thus the accretion could have occurred earlier, which would alleviate the problem 
with disk life-times. If so, stars with giant planets would be expected to 
show a chemical abundance pattern more similar to solar. 
Our sample of 10 solar analogs from planet surveys includes 
four solar analogs with giant planets, while for the other six 
no planets have been detected yet. 
As shown in Fig. 5 all solar analogs with giant planets differ from the Sun but closely 
resemble most solar twins. Thus, the odd solar composition is not due 
to giant planets as such. 
On the other hand, only two of the six solar analogs without close-in giant planets have 
abundances that differ significantly from the solar pattern. 
The fraction of stars with the solar pattern seems thus
tentatively related to the presence of giant planets on close orbits: $\approx 0$\% 
when having such planets, $\approx 20$\% for solar-type stars in general 
and $\approx 50-70$\% without close-in giant planets. 
No doubt, the statistics has to be improved considerably, 
%%MA
%before such numbers can be regarded reliable. 
but the numbers are clearly tantalizing. 

\section{Conclusions}
We conclude that Sun is unusual albeit not unique in its detailed elemental abundance pattern.
In particular it has a chemical composition that is 
more affected by dust condensation than most other
similar stars, such as most solar twins and 
all solar analogs with discovered giant planets.  
This may be purely circumstantial. For instance, the 
cloud in which the Sun 
was formed may first have been
cleansed from dust by radiation from hot luminous stars to a higher degree 
than happened where most other solar-type stars were formed. 
However, if an explanation of less {\it ad-hoc} character is searched for,
it is natural to explore whether the solar composition could be 
related to the way the formation of the planets in the solar system occurred.  
A reasonable hypothesis is then that the accretion onto the Sun
of the proto-planetary solar nebula, chemically affected by dust
condensation and planet  formation, was delayed long enough for
the cleansed gas to significantly affect the composition of the solar
convection zone.
%%MA
%\footnote{
%Another possibility recently suggested to by
%\citet{nordlund09}, and which follows from the results of the
%dynamical star-formation calculations of \cite{wuchterl03},
%and \cite{wuchterl01}, 
%is that the solar convection zone
%never included more than a minor fraction of a solar mass, in which
%case there is no need to assume a delayed accretion.
Another possibility is that the early Sun was never fully convective, which would 
make it easier to imprint a dust-cleansed abundance signature
\citep{nordlund09}. 
This scenario is supported by 
the dynamical star-formation calculations with a time-dependent convection treatment
of \cite{wuchterl03} and \cite{wuchterl01}.
%that imply that the proto-Sun was $\approx 500$\,K hotter and
%more luminous at a given age than pre-main sequence tracks assuming
%hydrostatic equilibrium \citep{dan94}.

The fact that stars with known close giant planets do not show the 
solar pattern suggests that their gas disks for some reason were accreted earlier 
when the stellar convection zones were so deep that the planetary signatures 
in the infalling gas disks were erased, and/or that the key factor behind the 
solar pattern is the formation and non-accretion of inner terrestrial planets. 
This would imply that solar-like stars with planetary systems similar to our own 
are a relatively rare occurrence. Before firm conclusions can be drawn regarding 
the effects and frequency of planet formation, accurate analyses of significantly 
larger samples of solar-like stars with and without known giant planets are important. 
It is reassuring that an independent analysis of 
22 northern solar twins confirm our findings both in terms of the magnitude
of the abundance difference between volatiles and refractories and the frequency
of stars showing the peculiar solar abundance pattern
\citep{ramirez09}. 
Our findings provide strong impetus to improve the abundance accuracy of 
late-type stars to the 0.01 dex level by removing the remaining systematic errors 
in the modelling of stellar atmospheres and line formation \citep{asp05, asp09}, 
which have obscured the 
planetary signatures in the stellar compositions until now. 
It is an enthralling prospect to be able to identify stars with planetary 
systems similar to the Solar system by means of their abundance patterns.

%% The \notetoeditor{TEXT} command allows the author to communicate
%% information to the copy editor.  This information will appear as a
%% footnote on the printed copy for the manuscript style file.  Nothing will
%% appear on the printed copy if the preprint or
%% preprint2 style files are used.

%% The eqnarray environment produces multi-line display math. The end of
%% each line is marked with a \\. Lines will be numbered unless the \\
%% is preceded by a \nonumber command.
%% Alignment points are marked by ampersands (&). There should be two
%% ampersands (&) per line.


%% Included in this acknowledgments section are examples of the
%% AASTeX hypertext markup commands. Use \url without the optional [HREF]
%% argument when you want to print the url directly in the text. Otherwise,
%% use either \url or \anchor, with the HREF as the first argument and the
%% text to be printed in the second.

\acknowledgments

Our results are based on observations made using Australian time at 
the Clay 6.5 m Magellan telescope at Las Campanas Observatory in Chile. 
Travel support was provided by the Australian Access to 
Major Facilities Programme (06/07-O-11). 
We thank A. Alves-Brito for help with the data reduction, J. Johnson
for her NH line list,
W. Lyra, M. Davies, T. Henning for discussions on disk lifetimes, 
planetary connections and effects of stellar passages, 
A. Serenelli, C. Straka, S. Vauclair and A. Weiss for discussions 
on stellar evolution and diffusion, and \AA{ke} Nordlund
for discussions on early stellar evolution and planetary formation.
%%MA
%and for his alternative explanation of the peculiar solar composition.

We acknowledge support from the 
Portuguese FCT (project PTDC/CTE-AST/65971/2006, and Ciencia 2007 program), 
the Australian Research Council, 
and the Swedish Research Council.


%% in parentheses, after the keyword, but they will not be verified.

{\it Facilities:} \facility{GMT (MIKE)}, \facility{Keck I (HIRES)}.

\clearpage

%% the file, uncomment the \plotone, \plottwo, and
%% If you need a layout that cannot be achieved with \plotone or
%% \plottwo, you can invoke the graphicx package directly with the
%% \includegraphics command or use \plotfiddle. For more information,
%% please see the tutorial on "Using Electronic Art with AASTeX" in the
%% documentation section at the AASTeX Web site,
%% http://www.journals.uchicago.edu/AAS/AASTeX.
%%
%% The examples below also include sample markup for submission of
%% supplemental electronic materials. As always, be sure to check
%% the instructions to authors for the journal you are submitting to
%% for specific submissions guidelines as they vary from

%% This example uses \plotone to include an EPS file scaled to
%% 80% of its natural size with \epsscale. Its caption
%% has been written to indicate that additional figure parts will be
%% available in the electronic journal.

\begin{figure}
\epsscale{.80}
\plotone{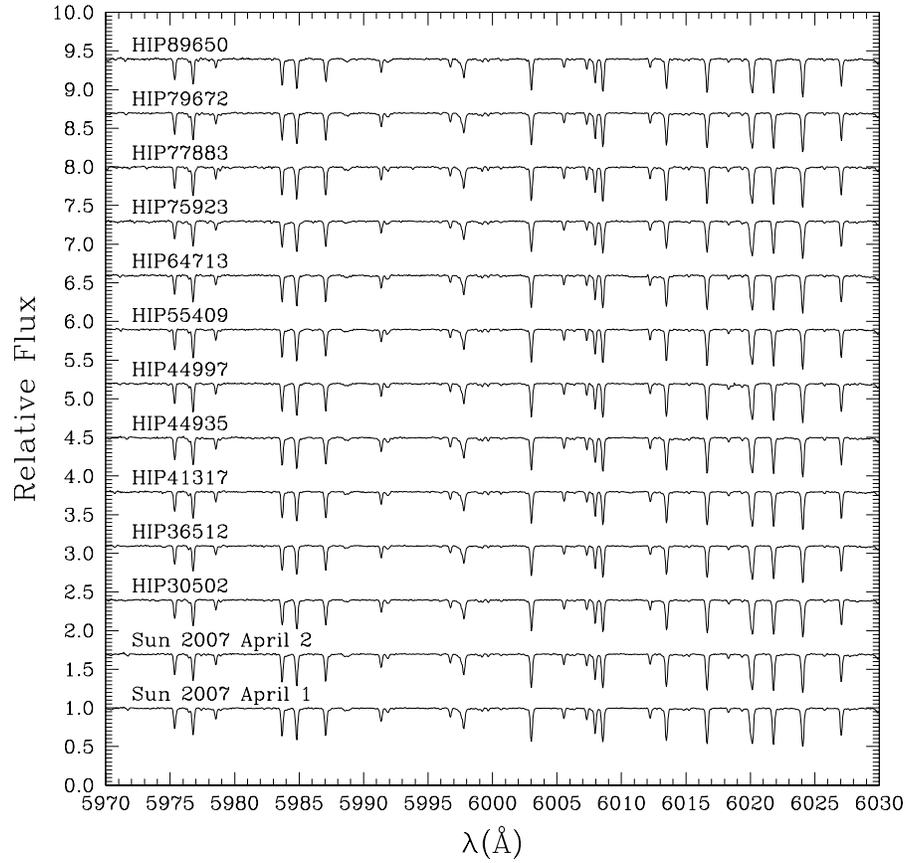}
\caption{High quality ($S/N ~\sim 450$) MIKE spectra of the 
Sun and solar twins in the region 5970-6030 \AA.
\label{fig1}}
\end{figure}

%\clearpage

\begin{figure}
\epsscale{.80}
\plotone{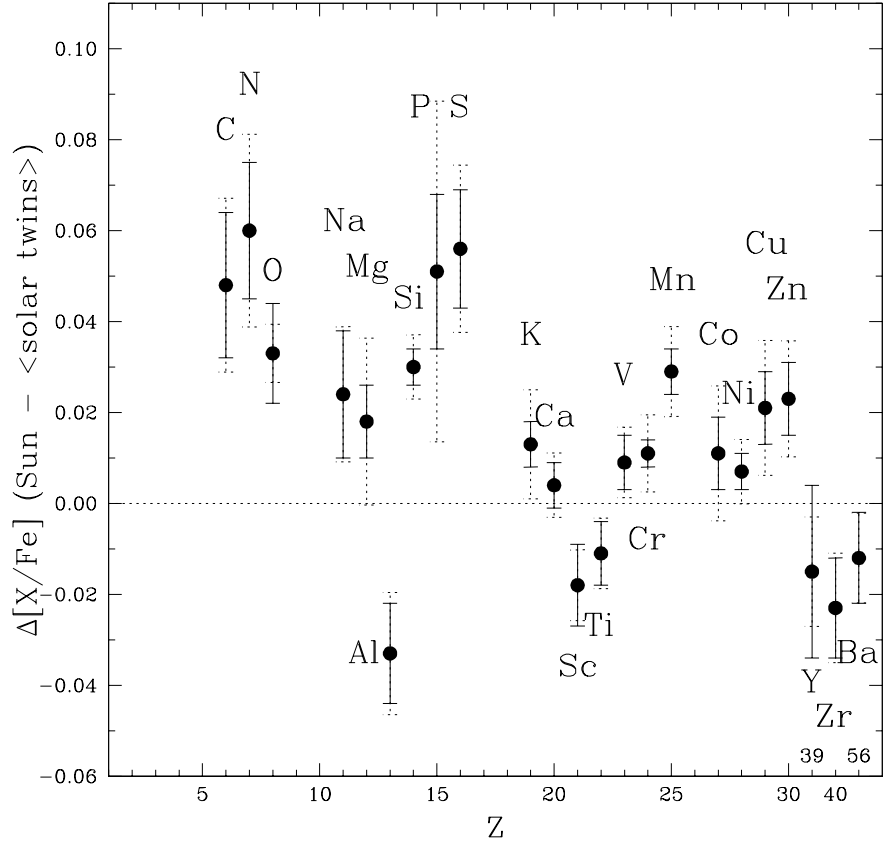}
\caption{Differences between [X/Fe] of the Sun and the 
mean values in the solar twins as a function 
of atomic number Z. For clarity the elements 
Y (Z = 39), Zr (Z = 40) and Ba (Z = 56) have been included after Zn.  
Observational $1\sigma$ errors in the relative abundances 
(including observational errors in both the Sun and solar twins) 
are shown with dotted error bars, while the $1\sigma$ errors in 
the mean abundance of the solar twins are shown with solid error bars.
\label{fig2}}
\end{figure}

%\clearpage

%% Here we use \plottwo to present two versions of the same figure,
%% one in black and white for print the other in RGB color
%% for online presentation. Note that the caption indicates
%% that a color version of the figure will be available online.
%%

\begin{figure}
\plotone{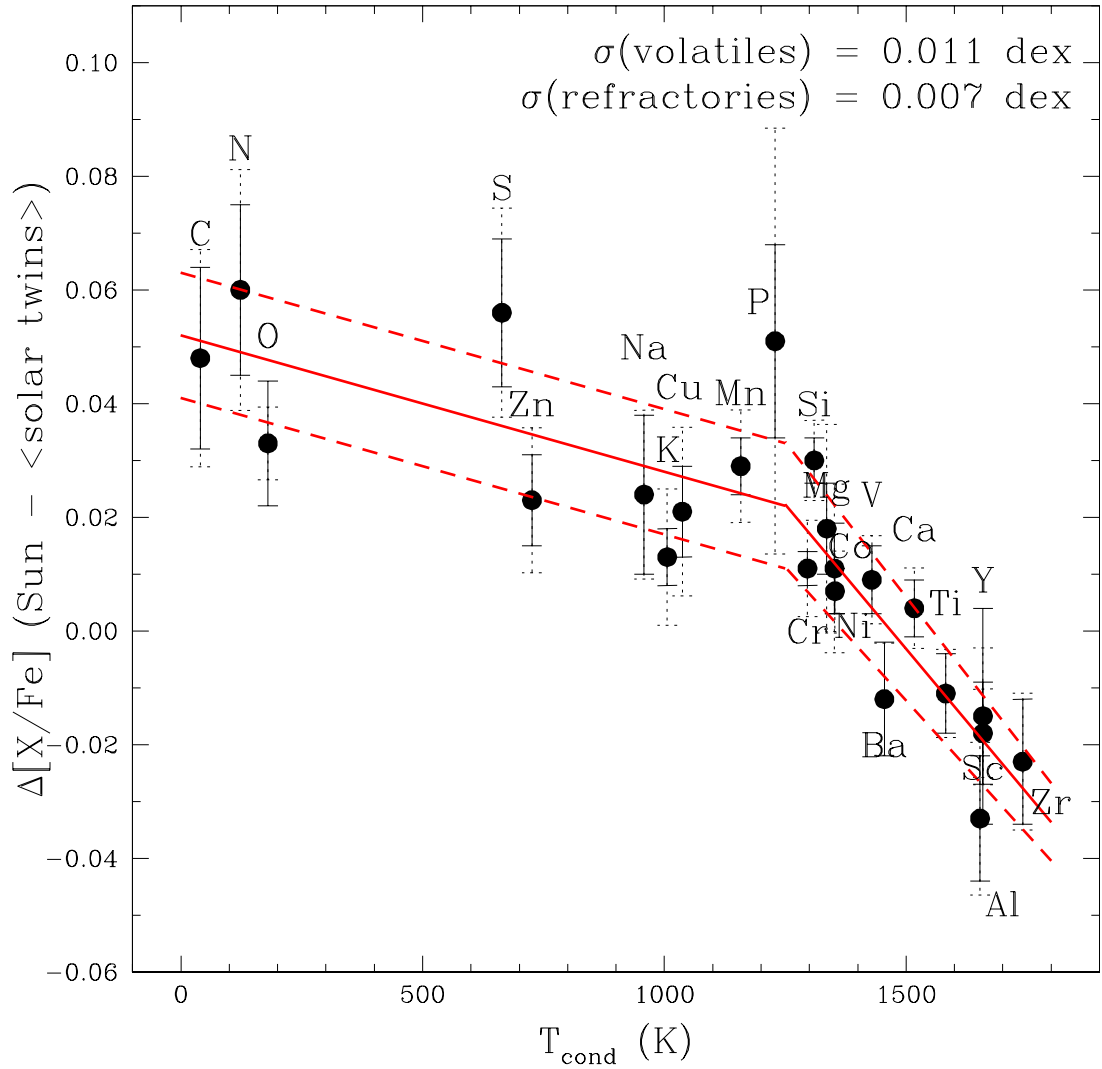}
\caption{Differences between [X/Fe] of the Sun and the 
mean values in the solar twins 
as a function of $T_{\rm cond}$. 
The abundance pattern shows a break at $T_{\rm cond} \sim 1200$ K. 
The solid lines are fits to the abundance pattern, while the dashed lines represent the 
standard deviation from the fits. 
The low element-to-element scatter from the 
fits for the refractory 
($\sigma = 0.007$ dex) and volatile ($\sigma = 0.011$ dex) elements 
confirms the high precision of our work.
The zero-point for the differences in relative chemical 
abundances depends on the adopted reference element, which here is Fe; 
the volatiles would appear normal while the refractories more depleted 
had we instead selected to use C. Error bars as in Fig. 2.
\label{fig3}}
\end{figure}

%% This figure uses \includegraphics to scale and rotate the still frame

\begin{figure}
\epsscale{.80}
\plotone{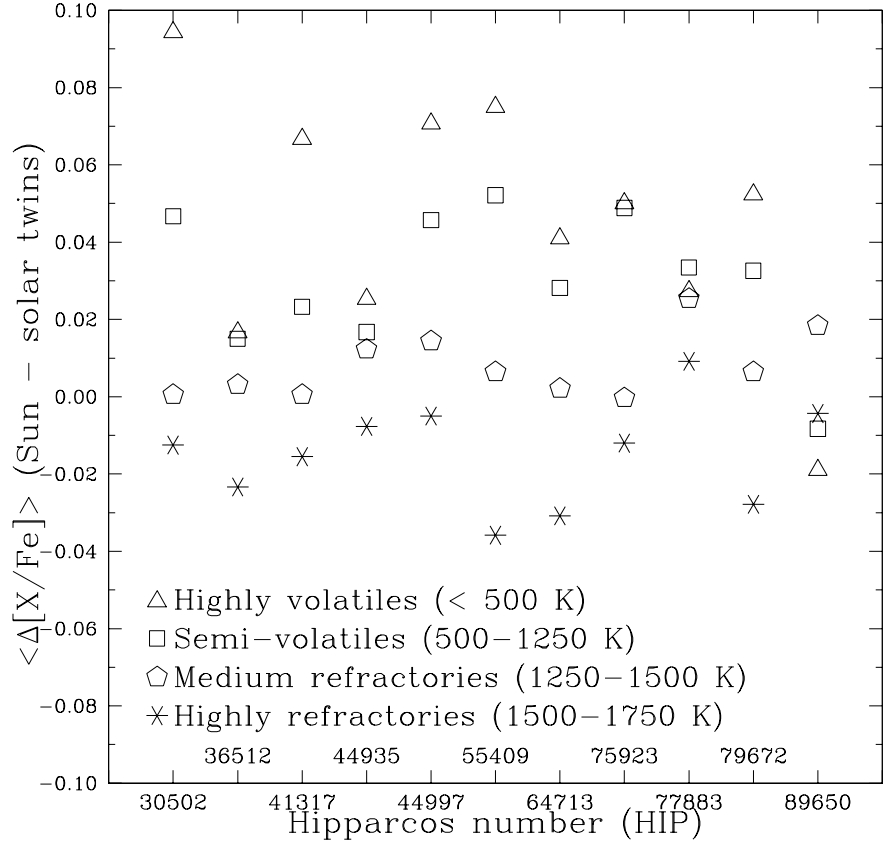}
\caption{Mean [X/Fe] ratios observed in the solar twins for highly volatile (triangles), 
semi-volatile (squares), medium refractory (pentagons) and 
highly refractory (stars) elements, for the solar twins. 
Most twins show similar trends with $T_{\rm cond}$ with the 
highly refractory elements ($1500<T_{\rm cond}<1750$ K) 
well separated from the highly volatile elements ($T_{\rm cond}<500$ K) 
and in between the semi-volatile ($500<T_{\rm cond}<1250$ K) and 
medium refractory ($1250<T_{\rm cond}<1500$ K) elements. 
This separation between refractories and volatiles on a 
star-by-star basis further emphasizes that the 
abundance differences are indeed real. 
\label{fig4} }
\end{figure}

\begin{figure}
\epsscale{.80}
\plotone{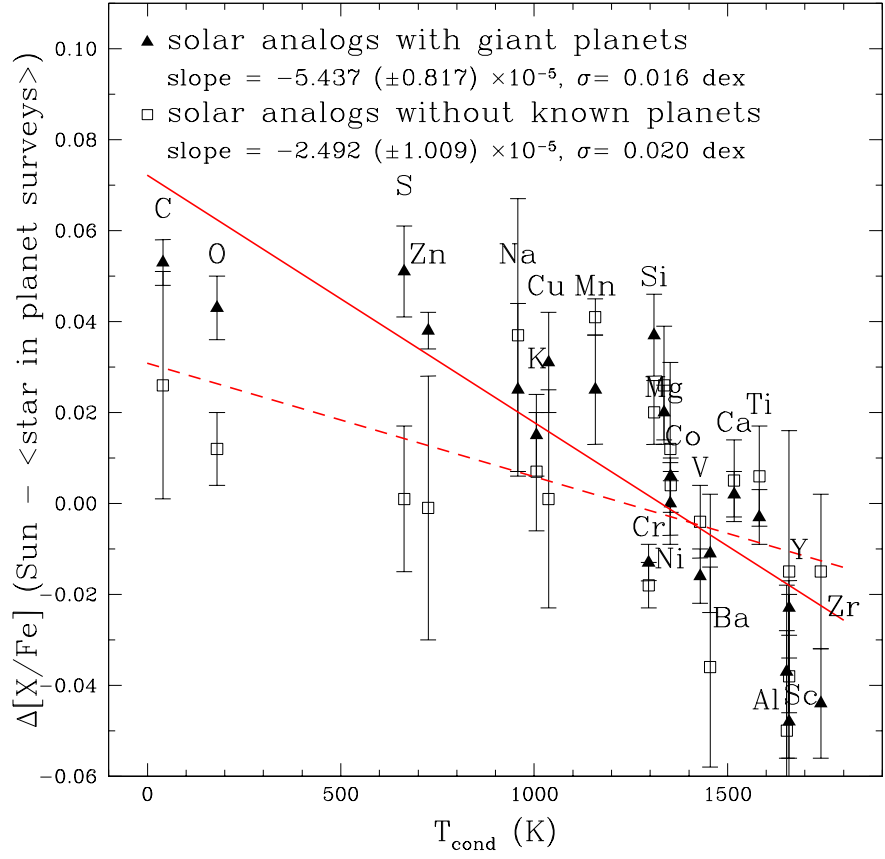}
\caption{Differences between [X/Fe] of the Sun and the 
mean values in the solar analogs with (filled triangles) and 
without (open squares) detected planets, as a function of their $T_{\rm cond}$. 
The planet hosting stars show a strong trend with $T_{\rm cond}$ 
almost identical to that seen in the solar twins (Fig. 3). 
The correlation is highly statistically significant 
($r_{\rm S} = -0.91$) with a probability of having this correlation by pure chance of $\sim10^{-8}$. 
Solar analogs without planets show a much shallower slope 
and a much weaker correlation. 
Linear fits to both samples are shown. 
The slope of the stars with planets is significant at the $7\sigma$ level, 
while the slope of the stars without planets is lower and significant only at the 
$2\sigma$ level.
\label{fig5}}
\end{figure}

%% If you are not including electonic art with your submission, you may
%% mark up your captions using the \figcaption command. See the
%% No more than seven \figcaption commands are allowed per page,
%% so if you have more than seven captions, insert a \clearpage

%% Tables should be submitted one per page, so put a \clearpage before

%% Two options are available to the author for producing tables:  the
%% deluxetable environment provided by the AASTeX package or the LaTeX
%% table environment.  Use of deluxetable is preferred.
%%

%% Three table samples follow, two marked up in the deluxetable environment,
%% one marked up as a LaTeX table.

%% In this first example, note that the \tabletypesize{}
%% command has been used to reduce the font size of the table.
%% We also use the \rotate command to rotate the table to
%% landscape orientation since it is very wide even at the
%% Note also that the \label command needs to be placed

%% This table also includes a table comment indicating that the full
%% version will be available in machine-readable format in the electronic

\end{document}